\documentclass[11pt]{article}

\usepackage[T1]{fontenc}
\usepackage[utf8]{inputenc}
\usepackage{lmodern}
\usepackage{microtype}

\usepackage[a4paper,margin=1in]{geometry}

\usepackage{amsmath,amssymb,mathtools,bm}

\usepackage{booktabs,tabularx,array}
\usepackage{xcolor}

\usepackage{cite}
\usepackage{hyperref}

\usepackage[mathlines,switch]{lineno}

\hypersetup{
  colorlinks=true,
  linkcolor=blue!55!black,
  citecolor=blue!55!black,
  urlcolor=blue!55!black,
  pdftitle={Gravitational Thermodynamicization of Primordial Black Holes},
  pdfauthor={Wen-Xiang Chen}
}

\allowdisplaybreaks
\newcommand{\dd}{\mathrm{d}}

\newcommand{\ii}{\mathrm{i}}
\newcommand{\Res}{\operatorname*{Res}}

\newcommand{\mpl}{M_{\mathrm{Pl}}}
\newcommand{\tpl}{t_{\mathrm{Pl}}}
\newcommand{\Tb}{T_{\mathrm{b}}}
\newcommand{\Trad}{T_{\mathrm{rad}}}
\newcommand{\SBH}{S_{\mathrm{BH}}}
\newcommand{\MH}{M_{\mathrm{H}}}
\newcommand{\Rh}{\mathcal{R}_{h}}
\newcommand{\Order}{\mathcal{O}}
\renewcommand{\TH}{T_{\mathrm H}}

\begin{document}

\title{
  \textbf{Gravitational Thermodynamicization of Primordial Black Holes:}\\
  An RVB--Residue Formulation of Thermal Formation and
  Nonequilibrium Evolution
}

\author{
  Wen-Xiang Chen\\
  School of Electronic Information\\
  Guangzhou City University of Technology\\
  \texttt{wxchen4277@qq.com}
}

\date{\today}

\maketitle


\begin{abstract}
We formulate a controlled thermodynamic description of neutral, nonrotating
primordial black holes (PBHs) by combining a local
Robson--Villari--Biancalana (RVB) residue prescription with the
thermodynamicization of gravitational dynamics.  For a nonextremal spherical
horizon of
\(
\dd s^2=-A(r)\dd t^2+\dd r^2/B(r)+r^2\dd\Omega_2^2
\),
the meromorphic one-form
\(
\omega_h=\dd r/\sqrt{A(r)B(r)}
\)
has a simple pole.  We show step by step that its positive residue
\(
\Rh=\left|\Res_{r=r_h}\omega_h\right|
\)
determines the inverse Hawking temperature,
\(
\beta_h=4\pi\Rh
\).
Along a one-parameter, fixed-coupling black-hole family, the first law then
gives the residue entropy
\(
S_{\rm res}(M)=S_{\rm ref}+4\pi\int_{M_{\rm ref}}^M\Rh(m)\dd m
\).
For the Schwarzschild branch, \(\Rh=2GM\), and the standard results
\(
\TH=(8\pi GM)^{-1}
\)
and
\(
\SBH=4\pi GM^2=A_h/(4G)
\)
are recovered with the entropy normalization \(S(0)=0\).

The resulting entropy defines an off-shell canonical exponent
\(
\mathcal{B}(M;\beta_b)=\beta_b M-S_{\rm res}(M)
\).
Its stationary point obeys \(\Tb=\TH\).  In the Schwarzschild case,
\(
M_\star=(8\pi G\Tb)^{-1}
\)
is an unstable thermal saddle and
\(
\mathcal{B}_\star=(16\pi G\Tb^2)^{-1}
\).
For a radiation-dominated Friedmann universe with constant effective number
of relativistic degrees of freedom \(g_\ast\), this becomes
\(
\mathcal{B}_\star(t)=\sqrt{\pi g_\ast/720}\,t/\tpl
\).
Consequently, equilibrium thermal production is exponentially suppressed
throughout the self-consistent semiclassical regime; this result does not
replace the usual density-perturbation collapse criterion.

For post-formation evolution we adopt a detailed-balance mass law
\(
\dot M=\mathcal{K}(M,t)(\Tb^4-\TH^4)
\)
with \(\mathcal{K}\geq0\), prove
\(
\dot S_{\rm gen}\geq0
\),
and express the dynamics as a free-energy gradient flow.  The moving mass
\(M_\star(t)\) is the instantaneous evaporation--accretion threshold;
the corresponding fixed-bath stationary point is linearly unstable.
Finally, for a small analytic deformation
\(
\Rh=2GM[1+\epsilon\psi(M)]
\),
we obtain explicit first-order shifts of the temperature, entropy, thermal
saddle, and, after fixing the entropy normalization, nucleation exponent.
The construction is local and
quasi-stationary; it is not asserted for degenerate horizons, horizon-scale
dynamics, or the Planck regime.
\end{abstract}


\section{Introduction}

Primordial black holes are distinguished from stellar remnants by their
possible formation before stars, through the collapse of sufficiently large
inhomogeneities in the early universe
\cite{Hawking1971,CarrHawking1974,Carr1975}.  Their possible contribution to
nonbaryonic dark matter was recognized at an early stage
\cite{Chapline1975}.  Modern calculations describe formation using
relativistic hydrodynamics, a profile-dependent threshold, and critical
collapse.  The relevant conceptual lineage runs from critical phenomena in
gravitational collapse
\cite{Choptuik1993,EvansColeman1994,KoikeHaraAdachi1995}
to PBH-specific mass scaling
\cite{NiemeyerJedamzik1998,NiemeyerJedamzik1999}, fully relativistic
simulations
\cite{MuscoMillerRezzolla2005}, and refined threshold prescriptions
\cite{HaradaYooKohri2013,Musco2019,MuscoEtAl2021}.

Black-hole mechanics and quantum field theory supply a second, logically
independent description.  The four laws of black-hole mechanics
\cite{BardeenCarterHawking1973}, Bekenstein entropy
\cite{Bekenstein1973}, and Hawking radiation
\cite{Hawking1975} establish that a stationary horizon has genuine
thermodynamic variables.  The generalized second law constrains exchanges
between the horizon and exterior matter
\cite{Bekenstein1974}.  The Unruh effect gives the corresponding local
observer-dependent thermal scale
\cite{Unruh1976}, while explicit greybody calculations determine the
species-dependent emission rate beyond the ideal blackbody approximation
\cite{Page1976,MacGibbon1991}.

These facts motivate the broader idea that gravitational field equations can
be read as thermodynamic equations of state or entropy-stationarity
conditions.  Jacobson derived the Einstein equation from local horizon heat
balance \cite{Jacobson1995}; Hayward formulated a unified first law in
spherical symmetry \cite{Hayward1998}; Padmanabhan developed horizon
thermodynamics and entropy-functional formulations
\cite{Padmanabhan2002,PadmanabhanParanjape2007,Padmanabhan2010}; and the
Friedmann equations themselves can be written as an apparent-horizon first
law \cite{CaiKim2005}.  Curvature corrections can require an explicitly
nonequilibrium entropy-production term \cite{ElingGuedensJacobson2006}.
The gravitational-thermodynamicization program considered by the present
author develops this entropy-functional viewpoint together with complex
analytic structures \cite{ChenGeometry2025}, while thermodynamic potentials
provide the appropriate fixed-temperature language
\cite{ChenThermoPotentials2026}.

The RVB method extracts Hawking temperature from the topology of a
dimensionally reduced Euclidean geometry \cite{RobsonVillariBiancalana2019}.
A residue-based extension has been applied to modified-gravity black holes
\cite{ChenCJP2025} and, in a local near-horizon form, to dynamical Vaidya and
Kinnersley geometries \cite{ChenVaidya2026}.  Complex paths also underlie
Hamilton--Jacobi and tunneling derivations of Hawking radiation
\cite{ParikhWilczek2000,AnghebenEtAl2005,HaywardEtAl2009}.  These approaches
suggest that the simple pole at a nonextremal horizon is not merely a
computational nuisance: its coefficient is the minimal local analytic datum
that controls thermal periodicity.

The purpose of this paper is therefore narrow.  We do not propose a new
collapse threshold and do not identify equilibrium thermal nucleation with
the standard PBH density-fluctuation channel.  Instead, we construct and test
the following map:
\begin{equation}
\boxed{
(A,B)
\xrightarrow{\ \Res\ }\Rh
\xrightarrow{\ \beta_h=4\pi\Rh\ }\TH
\xrightarrow{\ \dd S=\beta_h\dd M\ }S_{\rm res}
\xrightarrow{\ \mathcal{B}=\beta_bM-S_{\rm res}\ }
\{M_\star,\mathcal{B}_\star\}
\xrightarrow{\ \mathrm{balance}\ }
\{\dot M,\dot S_{\rm gen}\}.
}
\label{eq:central-map}
\end{equation}
The classical Schwarzschild limit reproduces the finite-temperature
instability found by Gross, Perry, and Yaffe
\cite{GrossPerryYaffe1982} and the Euclidean negative mode
\cite{Allen1984}.  The local advance is to make the residue the common
generator of the horizon temperature, entropy, canonical barrier, and
nonequilibrium PBH mass flow, and then to derive the radiation-era
consequences and the response to a general analytic residue deformation.

\section{Assumptions, conventions, and domain of validity}
\label{sec:assumptions}

We use \(c=\hbar=k_{\rm B}=1\), metric signature
\((-+++) \), and the unreduced Planck scales
\begin{equation}
\mpl=G^{-1/2},
\qquad
\tpl=\sqrt{G}.
\label{eq:planck-def}
\end{equation}
The dimensions in energy units are
\begin{equation}
[G]=E^{-2},\quad [M]=[T]=E,\quad [r]=[t]=E^{-1},
\quad [\Rh]=[\beta]=E^{-1},\quad [S]=[\mathcal{B}]=1.
\label{eq:dimensions}
\end{equation}
Thus every exponent used below is dimensionless.

The assumptions and conventions used throughout are summarized in
Table~\ref{tab:assumptions}.

\begin{table}[htbp]
\centering
\caption{Assumption and convention ledger.  The last column states what
would fail if the assumption were relaxed.}
\label{tab:assumptions}
\begin{tabularx}{\textwidth}{@{}p{0.19\textwidth}X p{0.27\textwidth}@{}}
\toprule
Item & Assumption used in this paper & Consequence or limitation \\
\midrule
Gravity
& Four-dimensional Einstein gravity is the classical baseline; possible
corrections enter only through a controlled deformation of the horizon
residue.
& A modified-gravity interpretation requires an independently solved metric
and its correct Wald entropy.\\

PBH state
& Neutral, nonrotating, approximately spherical PBH, parametrized by a single
mass \(M\).
& Spin, charge, chemical potentials, and varying couplings add work terms to
the first law.\\

Horizon
& The local horizon is a simple, nonextremal zero of the radial metric
functions.
& A double zero gives a leading double pole; the simple-pole temperature
prescription requires a separate extremal analysis.\\

Time dependence
& The PBH is subhorizon and quasi-stationary during one local thermal
timescale.
& The construction is not a global equilibrium temperature for a fully
dynamical cosmological spacetime.\\

Cosmology
& Spatially flat, radiation-dominated FRW background; \(g_\ast\) is constant
over the interval being differentiated.
& Threshold crossings of particle species require piecewise evolution of
\(g_\ast(T)\).\\

Reservoir
& The ambient radiation is locally thermal with temperature \(\Tb(t)\), and
the exchange kernel obeys detailed balance.
& A nonthermal distribution requires a spectral transport integral rather
than a difference of fourth powers.\\

Semiclassical regime
& \(M\gg\mpl\), curvature at the horizon is sub-Planckian, and backreaction is
encoded only through \(M(t)\).
& No statement is made about the final Planckian evaporation stage or a
quantum-gravity remnant.\\
\bottomrule
\end{tabularx}
\end{table}

The word ``thermodynamicization'' is used operationally: geometric horizon
data are mapped into thermodynamic state functions and transport laws.  We
do not assume that a thermodynamic analogy alone proves microscopic
emergence of spacetime.

\section{Radiation-era and PBH formation anchors}
\label{sec:frw}

\subsection{Radiation-dominated Friedmann solution}

For a spatially flat Friedmann--Robertson--Walker geometry,
\begin{equation}
\dd s^2=-\dd t^2+a^2(t)\dd\bm{x}^2,
\label{eq:frw-metric}
\end{equation}
the Friedmann and continuity equations are
\begin{align}
H^2&=\left(\frac{\dot a}{a}\right)^2=\frac{8\pi G}{3}\rho,
\label{eq:friedmann}\\
\dot\rho+3H(\rho+p)&=0.
\label{eq:continuity}
\end{align}
For radiation, \(p=\rho/3\).  Equation \eqref{eq:continuity} becomes
\begin{equation}
\dot\rho+4\frac{\dot a}{a}\rho=0.
\label{eq:rad-cont}
\end{equation}
Dividing by \(\rho\) and integrating gives
\begin{align}
\frac{\dd\rho}{\rho}&=-4\frac{\dd a}{a},
\nonumber\\
\ln\rho&=-4\ln a+\mathrm{constant},
\nonumber\\
\rho&=\rho_0\left(\frac{a_0}{a}\right)^4.
\label{eq:rho-a}
\end{align}
Substitution in \eqref{eq:friedmann} yields
\(\dot a\propto a^{-1}\), hence
\begin{equation}
a(t)\propto t^{1/2},
\qquad
H(t)=\frac{1}{2t}.
\label{eq:a-H-rad}
\end{equation}
Putting this result back into \eqref{eq:friedmann},
\begin{equation}
\rho(t)=\frac{3H^2}{8\pi G}
=\frac{3}{32\pi Gt^2}.
\label{eq:rho-t}
\end{equation}

For a thermal bath with effective energy degrees of freedom \(g_\ast\),
\begin{equation}
\rho_{\rm rad}=\frac{\pi^2}{30}g_\ast\Trad^4.
\label{eq:rho-T}
\end{equation}
Equating \eqref{eq:rho-t} and \eqref{eq:rho-T} gives, without skipping the
algebra,
\begin{align}
\frac{\pi^2}{30}g_\ast\Trad^4
&=\frac{3}{32\pi Gt^2},
\nonumber\\
\Trad^4(t)
&=\frac{90}{32\pi^3Gg_\ast t^2}
=\frac{45}{16\pi^3Gg_\ast t^2},
\label{eq:Trad4}\\
\Trad(t)
&=\left(\frac{45}{16\pi^3Gg_\ast}\right)^{1/4}t^{-1/2}.
\label{eq:Trad}
\end{align}

The mass inside one Hubble sphere, \(R_H=H^{-1}\), is
\begin{align}
\MH(t)
&=\frac{4\pi}{3}\rho H^{-3}
\nonumber\\
&=\frac{4\pi}{3}
\left(\frac{3H^2}{8\pi G}\right)H^{-3}
\nonumber\\
&=\frac{1}{2GH}
=\frac{t}{G}.
\label{eq:horizon-mass}
\end{align}
This establishes the standard order-of-magnitude relation between PBH mass
and formation time used in early PBH work
\cite{CarrHawking1974,Carr1975}.

\subsection{Density-perturbation channel}

Let \(\delta\) be a suitably smoothed density or compaction perturbation at
horizon entry.  Numerical studies show that collapse requires
\begin{equation}
\delta>\delta_c,
\label{eq:delta-threshold}
\end{equation}
where \(\delta_c\) depends on the perturbation profile and slicing.  Near
threshold, the PBH mass obeys the critical law
\begin{equation}
M_i=K\,\MH(t_i)\bigl(\delta-\delta_c\bigr)^{\gamma_c},
\qquad \delta>\delta_c,
\label{eq:critical-scaling}
\end{equation}
with a radiation-fluid exponent near \(\gamma_c\simeq0.36\), while \(K\) is
profile dependent
\cite{EvansColeman1994,KoikeHaraAdachi1995,NiemeyerJedamzik1998,
NiemeyerJedamzik1999,MuscoMillerRezzolla2005}.

For an illustrative Gaussian perturbation distribution with variance
\(\sigma^2(M_H)\), a Press--Schechter-type tail estimate
\cite{PressSchechter1974} is
\begin{equation}
\beta_{\rm coll}(M_H)
\simeq
\int_{\delta_c}^{\infty}
\frac{\dd\delta}{\sqrt{2\pi}\sigma}
\exp\left[-\frac{\delta^2}{2\sigma^2}\right].
\label{eq:gaussian-tail}
\end{equation}
When \(\nu_c\equiv\delta_c/\sigma\gg1\), integration by parts or the
asymptotic complementary-error-function expansion gives
\begin{equation}
\beta_{\rm coll}(M_H)
\simeq
\frac{\sigma}{\sqrt{2\pi}\delta_c}
\exp\left[-\frac{\delta_c^2}{2\sigma^2}\right]
\left[1+\Order\left(\frac{\sigma^2}{\delta_c^2}\right)\right].
\label{eq:tail-asymptotic}
\end{equation}
Equations \eqref{eq:delta-threshold}--\eqref{eq:tail-asymptotic} describe a
nonequilibrium gravitational-collapse channel.  The canonical thermal
exponent derived in Sec.~\ref{sec:nucleation} is a different quantity and
must not be substituted for \(\delta_c^2/(2\sigma^2)\).

\section{RVB--residue temperature from a simple horizon}
\label{sec:residue-temp}

\subsection{General two-function spherical metric}

Consider the static spherical line element
\begin{equation}
\dd s^2=-A(r)\dd t^2+\frac{\dd r^2}{B(r)}
+r^2\dd\Omega_2^2.
\label{eq:ABmetric}
\end{equation}
Let \(r=r_h\) be a nonextremal horizon:
\begin{equation}
A(r_h)=B(r_h)=0,
\qquad
a_1\equiv A'(r_h)\neq0,
\qquad
b_1\equiv B'(r_h)\neq0.
\label{eq:simple-horizon}
\end{equation}
Writing \(z=r-r_h\), Taylor expansion gives
\begin{align}
A(r)&=a_1z+\frac{a_2}{2}z^2+\Order(z^3),
\label{eq:Aexp}\\
B(r)&=b_1z+\frac{b_2}{2}z^2+\Order(z^3).
\label{eq:Bexp}
\end{align}
The product is
\begin{equation}
A(r)B(r)=a_1b_1z^2+
\left(\frac{a_1b_2+a_2b_1}{2}\right)z^3+\Order(z^4),
\label{eq:ABexp}
\end{equation}
and therefore
\begin{equation}
\frac{1}{\sqrt{A(r)B(r)}}
=\frac{1}{\sqrt{a_1b_1}}\frac{1}{z}
+\Order(z^0).
\label{eq:inverseABexp}
\end{equation}
The horizon one-form
\begin{equation}
\omega_h\equiv\frac{\dd r}{\sqrt{A(r)B(r)}}
\label{eq:horizon-oneform}
\end{equation}
thus has a simple pole with positive physical residue
\begin{equation}
\Rh\equiv
\left|\Res_{r=r_h}\omega_h\right|
=\frac{1}{\sqrt{|a_1b_1|}}.
\label{eq:general-residue}
\end{equation}

\subsection{Euclidean regularity, shown explicitly}

After \(t=-\ii\tau\), the near-horizon Euclidean two-metric is
\begin{equation}
\dd s_E^2
\simeq a_1z\,\dd\tau^2+\frac{\dd z^2}{b_1z}.
\label{eq:euclidean-near}
\end{equation}
Define a proper radial coordinate
\begin{equation}
\rho\equiv2\sqrt{\frac{z}{b_1}},
\qquad
z=\frac{b_1}{4}\rho^2,
\qquad
\dd z=\frac{b_1}{2}\rho\,\dd\rho.
\label{eq:rho-def}
\end{equation}
The radial term becomes
\begin{equation}
\frac{\dd z^2}{b_1z}
=
\frac{(b_1^2\rho^2/4)\dd\rho^2}
{b_1(b_1\rho^2/4)}
=\dd\rho^2,
\label{eq:radial-flat}
\end{equation}
while the Euclidean-time term becomes
\begin{equation}
a_1z\,\dd\tau^2
=\frac{a_1b_1}{4}\rho^2\dd\tau^2.
\label{eq:time-polar}
\end{equation}
Consequently,
\begin{equation}
\dd s_E^2\simeq
\dd\rho^2+\kappa_h^2\rho^2\dd\tau^2,
\qquad
\kappa_h\equiv\frac{1}{2}\sqrt{|a_1b_1|}.
\label{eq:polar-metric}
\end{equation}
The origin \(\rho=0\) is regular only if
\(\theta=\kappa_h\tau\) has period \(2\pi\).  Hence
\begin{align}
\beta_h
&=\frac{2\pi}{|\kappa_h|}
=\frac{4\pi}{\sqrt{|a_1b_1|}}
=4\pi\Rh,
\label{eq:beta-residue}\\
\TH
&=\beta_h^{-1}
=\frac{1}{4\pi\Rh}
=\frac{\sqrt{|a_1b_1|}}{4\pi}.
\label{eq:T-residue}
\end{align}
Equation \eqref{eq:T-residue} is the central local RVB--residue
temperature formula.

\subsection{Contour form and analytic coordinate invariance}

By the residue theorem,
\begin{equation}
\oint_{\mathcal C_h}\omega_h
=2\pi\ii\Res_{r=r_h}\omega_h,
\label{eq:contour}
\end{equation}
where \(\mathcal C_h\) is a sufficiently small positively oriented contour
around the complexified horizon.  Therefore, up to the orientation sign,
\begin{equation}
\beta_h
=4\pi
\left|\frac{1}{2\pi\ii}
\oint_{\mathcal C_h}\frac{\dd r}{\sqrt{A(r)B(r)}}\right|.
\label{eq:beta-contour}
\end{equation}

Let \(u=u(r)\) be analytic near the horizon with
\(\dd u/\dd r\neq0\).  Since
\begin{equation}
\frac{\dd r^2}{B(r)}
=\frac{\dd u^2}{\widetilde B(u)},
\qquad
\widetilde B(u)
=B(r(u))\left(\frac{\dd u}{\dd r}\right)^2,
\label{eq:Btransform}
\end{equation}
we obtain
\begin{equation}
\frac{\dd u}{\sqrt{A(r(u))\widetilde B(u)}}
=\frac{\dd r}{\sqrt{A(r)B(r)}}
\label{eq:oneform-invariance}
\end{equation}
for an orientation-preserving branch.  Thus the residue is invariant as the
coefficient of a meromorphic one-form, provided that the normalization of
the physical time coordinate is held fixed.  A rescaling of \(t\) changes
the measured temperature and is not a gauge-invariant radial
reparametrization.

\subsection{Relation to the original RVB topological formula}

For the commonly used gauge \(A=B=F\), the Euclidean two-metric is
\begin{equation}
\dd s_{E,2}^2=F(r)\dd\tau^2+\frac{\dd r^2}{F(r)},
\qquad
\sqrt{g_2}=1,
\qquad
R^{(2)}=-F''(r).
\label{eq:2d-curvature}
\end{equation}
The Gauss--Bonnet theorem for a disk-like Euclidean section is
\begin{equation}
\chi=
\frac{1}{4\pi}\int_{\mathcal M}R^{(2)}\sqrt{g_2}\,\dd^2x
+\frac{1}{2\pi}\int_{\partial\mathcal M}k_g\,\dd s.
\label{eq:GB}
\end{equation}
Regulate the disk by \(r_h+\varepsilon\leq r\leq R\).  The bulk term is
\begin{align}
\frac{1}{4\pi}\int_0^\beta\dd\tau
\int_{r_h+\varepsilon}^{R}[-F''(r)]\dd r
&=
\frac{\beta}{4\pi}
\left[F'(r_h+\varepsilon)-F'(R)\right].
\label{eq:GBbulk}
\end{align}
The outer-boundary geodesic-curvature term supplies
\(\beta F'(R)/(4\pi)\).  Taking \(\varepsilon\to0\), filling in the regular
horizon cap, and using \(\chi=1\), Eq.~\eqref{eq:GB} becomes
\begin{equation}
1=\frac{\beta F'(r_h)}{4\pi},
\qquad
\beta=\frac{4\pi}{F'(r_h)}.
\label{eq:GBbeta}
\end{equation}
Since
\begin{equation}
\Res_{r=r_h}\frac{\dd r}{F(r)}
=\frac{1}{F'(r_h)},
\label{eq:Fres}
\end{equation}
the local residue formula \eqref{eq:beta-residue} is precisely the
simple-horizon, one-disk realization of the RVB topological construction
\cite{RobsonVillariBiancalana2019}.  The topological formula is global,
whereas Eq.~\eqref{eq:beta-contour} deliberately isolates one local horizon.

\subsection{Schwarzschild and slowly evolving Vaidya PBHs}

For a Schwarzschild PBH,
\begin{equation}
A(r)=B(r)=F(r)=1-\frac{2GM}{r},
\qquad
r_h=2GM.
\label{eq:schwarzschild-F}
\end{equation}
Differentiating,
\begin{equation}
F'(r)=\frac{2GM}{r^2},
\qquad
F'(r_h)
=\frac{2GM}{(2GM)^2}
=\frac{1}{2GM}.
\label{eq:Fprime-sch}
\end{equation}
Therefore,
\begin{equation}
\Rh=\frac{1}{F'(r_h)}=2GM,
\qquad
\beta_h=8\pi GM,
\qquad
\TH=\frac{1}{8\pi GM},
\label{eq:sch-temp}
\end{equation}
in agreement with Hawking's result \cite{Hawking1975}.

A slowly accreting or evaporating spherical PBH can be represented locally
by the Vaidya form
\begin{equation}
\dd s^2
=-F(v,r)\dd v^2+2\dd v\dd r+r^2\dd\Omega_2^2,
\qquad
F(v,r)=1-\frac{2GM(v)}{r}.
\label{eq:vaidya}
\end{equation}
At fixed advanced time \(v\), the inverse horizon function again has
\begin{equation}
\Res_{r=2GM(v)}\frac{\dd r}{F(v,r)}=2GM(v),
\label{eq:vaidya-res}
\end{equation}
and hence
\begin{equation}
\TH(v)=\frac{1}{8\pi GM(v)}.
\label{eq:vaidya-temp}
\end{equation}
This is a local trapping-horizon temperature, not a global equilibrium
temperature \cite{HaywardEtAl2009,ChenVaidya2026}.  Two useful adiabaticity
parameters are
\begin{equation}
\varepsilon_H\equiv2GMH\ll1,
\qquad
\varepsilon_{\rm ad}\equiv
\frac{|\dot\kappa_h|}{\kappa_h^2}
=4G|\dot M|\ll1,
\label{eq:adiabaticity}
\end{equation}
where \(\kappa_h=1/(4GM)\) was used in the final equality.

\section{Residue entropy and gravitational thermodynamicization}
\label{sec:entropy}

\subsection{Entropy reconstructed from the first law}

Along a one-parameter family with fixed charge, angular momentum, and
couplings, the first law reduces to
\begin{equation}
\dd M=\TH\,\dd S.
\label{eq:firstlaw-simple}
\end{equation}
Using \(\TH^{-1}=4\pi\Rh\),
\begin{equation}
\frac{\dd S_{\rm res}}{\dd M}
=\frac{1}{\TH}
=4\pi\Rh(M).
\label{eq:dSdM-res}
\end{equation}
Integrating from a reference mass \(M_{\rm ref}\),
\begin{equation}
S_{\rm res}(M)
=S_{\rm ref}
+4\pi\int_{M_{\rm ref}}^{M}\Rh(m)\dd m.
\label{eq:res-entropy-general}
\end{equation}
This equation is a reconstruction along a fixed thermodynamic path.  If
additional state variables vary, the correct law is instead
\begin{equation}
\dd M=T\dd S+\Omega\dd J+\Phi\dd Q+\sum_i\mathcal X_i\dd\lambda_i,
\label{eq:firstlaw-work}
\end{equation}
and Eq.~\eqref{eq:res-entropy-general} cannot be used without holding
\((J,Q,\lambda_i)\) fixed.  In a general diffeomorphism-invariant theory,
the resulting entropy must also agree with the appropriate Noether-charge
entropy \cite{Wald1993}.

For Schwarzschild, choose \(S(0)=0\) within the formal classical
extrapolation.  Substitution of \(\Rh=2GM\) gives
\begin{align}
\SBH(M)
&=4\pi\int_0^M2Gm\,\dd m
\nonumber\\
&=8\pi G\left[\frac{m^2}{2}\right]_0^M
\nonumber\\
&=4\pi GM^2.
\label{eq:S-sch}
\end{align}
The horizon area is
\begin{equation}
A_h=4\pi r_h^2
=4\pi(2GM)^2
=16\pi G^2M^2,
\label{eq:Ah-sch}
\end{equation}
so
\begin{equation}
\frac{A_h}{4G}=4\pi GM^2=\SBH,
\label{eq:area-law}
\end{equation}
recovering the Bekenstein--Hawking law
\cite{Bekenstein1973,Hawking1975}.

\subsection{Heat capacity directly from the residue}

From
\begin{equation}
T(M)=\frac{1}{4\pi\Rh(M)},
\label{eq:T-R}
\end{equation}
differentiation gives
\begin{equation}
\frac{\dd T}{\dd M}
=-\frac{\Rh'(M)}{4\pi\Rh^2(M)}.
\label{eq:dTdM}
\end{equation}
Therefore the fixed-work-variable heat capacity is
\begin{equation}
C\equiv\frac{\dd M}{\dd T}
=-\frac{4\pi\Rh^2(M)}{\Rh'(M)}.
\label{eq:C-res}
\end{equation}
For \(\Rh=2GM\),
\begin{equation}
C_{\rm Sch}
=-\frac{4\pi(2GM)^2}{2G}
=-8\pi GM^2<0.
\label{eq:C-sch}
\end{equation}
Thus the sign of \(\Rh'\) is a compact stability diagnostic: a growing
residue implies negative canonical heat capacity.

\section{Canonical thermal saddle in a radiation bath}
\label{sec:nucleation}

\subsection{Off-shell exponent generated by the residue}

Let the bath inverse temperature be
\(\beta_b=1/\Tb\).  The leading canonical weight of a configuration with
mass \(M\) and entropy \(S_{\rm res}(M)\) is
\begin{equation}
\exp[-\mathcal B(M;\beta_b)],
\qquad
\mathcal B(M;\beta_b)
\equiv\beta_bM-S_{\rm res}(M).
\label{eq:Bdef}
\end{equation}
At a stationary point,
\begin{align}
0
&=\left(\frac{\partial\mathcal B}{\partial M}\right)_{\beta_b}
\nonumber\\
&=\beta_b-\frac{\dd S_{\rm res}}{\dd M}
\nonumber\\
&=\beta_b-4\pi\Rh(M).
\label{eq:Bstationarity}
\end{align}
Hence
\begin{equation}
\beta_b=\beta_h(M_\star),
\qquad
\Tb=\TH(M_\star).
\label{eq:temp-match}
\end{equation}
The second derivative is
\begin{equation}
\left.
\frac{\partial^2\mathcal B}{\partial M^2}
\right|_{M_\star}
=-4\pi\Rh'(M_\star).
\label{eq:Bsecond}
\end{equation}
Combining Eqs.~\eqref{eq:C-res} and \eqref{eq:Bsecond}, the sign of the
canonical saddle curvature equals the sign of the heat capacity.

For Schwarzschild,
\begin{equation}
\mathcal B_{\rm Sch}(M;\beta_b)
=\beta_bM-4\pi GM^2.
\label{eq:Bsch}
\end{equation}
The first derivative is
\begin{equation}
\frac{\partial\mathcal B_{\rm Sch}}{\partial M}
=\beta_b-8\pi GM.
\label{eq:Bschprime}
\end{equation}
Solving \(\partial_M\mathcal B_{\rm Sch}=0\) gives
\begin{equation}
M_\star
=\frac{\beta_b}{8\pi G}
=\frac{1}{8\pi G\Tb}.
\label{eq:Mstar}
\end{equation}
The curvature is
\begin{equation}
\mathcal B_{\rm Sch}''(M_\star)=-8\pi G<0,
\label{eq:Bnegative}
\end{equation}
so the stationary black hole is the top of a barrier, not a stable
canonical equilibrium.  Substituting Eq.~\eqref{eq:Mstar} into
Eq.~\eqref{eq:Bsch},
\begin{align}
\mathcal B_\star
&=\beta_b\left(\frac{\beta_b}{8\pi G}\right)
-4\pi G\left(\frac{\beta_b}{8\pi G}\right)^2
\nonumber\\
&=\frac{\beta_b^2}{8\pi G}
-\frac{\beta_b^2}{16\pi G}
\nonumber\\
&=\frac{\beta_b^2}{16\pi G}
=\frac{1}{16\pi G\Tb^2}.
\label{eq:Bstar}
\end{align}
At the saddle,
\begin{equation}
\mathcal B_\star
=4\pi GM_\star^2
=\SBH(M_\star),
\qquad
F_\star\equiv\frac{\mathcal B_\star}{\beta_b}
=\frac{M_\star}{2}.
\label{eq:Bstar-entropy}
\end{equation}
The free energy \(F=M-\Tb S\) returns to zero at \(M=2M_\star\), after
which the ideal infinite-reservoir canonical description runs away.  This
is the thermodynamic manifestation of the finite-temperature instability
of asymptotically flat space \cite{GibbonsHawking1977,GrossPerryYaffe1982}.

\subsection{Radiation-era exponent and critical mass}

Set \(\Tb=\Trad(t)\) from Eq.~\eqref{eq:Trad}.  The instantaneous thermal
saddle is
\begin{align}
M_\star(t)
&=\frac{1}{8\pi G}
\left(\frac{16\pi^3Gg_\ast}{45}\right)^{1/4}t^{1/2}.
\label{eq:Mstar-t}
\end{align}
Dividing by the Hubble mass \(\MH=t/G\),
\begin{align}
\frac{M_\star(t)}{\MH(t)}
&=
\frac{1}{8\pi\Trad(t)t}
\nonumber\\
&=
\frac{1}{8\pi}
\left(\frac{16\pi^3g_\ast}{45}\right)^{1/4}
\left(\frac{\tpl}{t}\right)^{1/2}.
\label{eq:Mstar-MH}
\end{align}
For fixed \(g_\ast\), the ratio \(M_\star/\MH\) tends to zero as
\(t/\tpl\) grows.  More precisely, setting
\(x=t/\tpl\) and \(C_\ast=\sqrt{g_\ast}/(16\sqrt{45\pi})\) gives
\((M_\star/\MH)^2=C_\ast/x\) and
\((M_\star/\mpl)^2=C_\ast x\).
The subhorizon and semiclassical conditions therefore require
\(x\gg\max(C_\ast,C_\ast^{-1})\).  In this regime the local metric
approximation is supported, but the nucleation exponent is simultaneously
large.

To see this explicitly, use Eq.~\eqref{eq:Trad4}:
\begin{align}
\mathcal B_\star(t)
&=\frac{1}{16\pi G\Trad^2(t)}
\nonumber\\
&=\frac{1}{16\pi G}
\left(\frac{16\pi^3Gg_\ast t^2}{45}\right)^{1/2}
\nonumber\\
&=\frac{4\pi^{3/2}\sqrt{Gg_\ast}\,t}
{16\pi G\sqrt{45}}
\nonumber\\
&=\frac{\sqrt{\pi g_\ast}}{4\sqrt{45}}
\frac{t}{\sqrt G}
\nonumber\\
&=\sqrt{\frac{\pi g_\ast}{720}}\,
\frac{t}{\tpl}.
\label{eq:Bstar-t}
\end{align}

A thermal nucleation rate has the semiclassical form
\begin{equation}
\frac{\Gamma_{\rm th}}{V}
=\mathcal P(\Tb)\exp[-\mathcal B_\star],
\label{eq:nucleation-rate}
\end{equation}
where \(\mathcal P\) is a fluctuation prefactor not determined by the present
one-dimensional thermodynamic reduction.  The exponential conclusion is
nevertheless robust.  Indeed, from Eq.~\eqref{eq:Bstar-entropy},
\begin{equation}
\mathcal B_\star
=4\pi\left(\frac{M_\star}{\mpl}\right)^2.
\label{eq:Bstar-semiclassical}
\end{equation}
The validity condition \(M_\star\gg\mpl\) therefore implies
\(\mathcal B_\star\gg1\).  Equilibrium thermal formation of semiclassical
PBHs is exponentially rare.  Near \(t\sim\tpl\), where
Eq.~\eqref{eq:Bstar-t} need not be large, the semiclassical derivation is
not trustworthy.  This closes an important logical loophole: the
thermodynamic barrier is predictive precisely where it is suppressive, and
it cannot be extrapolated to claim unsuppressed Planckian production.

\subsection{Why this is not the usual PBH abundance}

The exponent in Eq.~\eqref{eq:nucleation-rate} describes an approximately
equilibrium Schwarzschild saddle in a local radiation reservoir.  The
standard PBH channel instead follows the nonlinear evolution of a
superhorizon curvature perturbation through horizon entry and is governed
by Eqs.~\eqref{eq:delta-threshold}--\eqref{eq:tail-asymptotic}.  Therefore
\begin{equation}
\boxed{
\mathcal B_\star
\neq
\frac{\delta_c^2}{2\sigma^2}
\quad\text{in general}.
}
\label{eq:channels-distinct}
\end{equation}
The residue formalism can supply the thermodynamics of a PBH after a local
horizon exists; it does not, by itself, calculate the primordial curvature
power spectrum, the compaction profile, or \(\delta_c\).

\section{Nonequilibrium PBH--radiation exchange}
\label{sec:noneq}

\subsection{Generalized entropy production}

Let \(\dot M>0\) mean net energy flow from the bath into the PBH.  Over a
local exchange interval short compared with the Hubble time,
\begin{equation}
\dot S_{\rm BH}=\frac{\dot M}{\TH},
\qquad
\dot S_{\rm bath}=-\frac{\dot M}{\Tb}.
\label{eq:entropy-rates}
\end{equation}
The exchange contribution to generalized entropy production is
\begin{equation}
\dot S_{\rm gen}
=\dot M\left(\frac{1}{\TH}-\frac{1}{\Tb}\right).
\label{eq:Sgen-basic}
\end{equation}
Assume a greybody-averaged detailed-balance law
\begin{equation}
\dot M
=\mathcal K(M,t)\left(\Tb^4-\TH^4\right),
\qquad
\mathcal K(M,t)\geq0.
\label{eq:mass-balance}
\end{equation}
Factorization gives
\begin{equation}
\Tb^4-\TH^4
=(\Tb-\TH)
\left(\Tb^3+\Tb^2\TH+\Tb\TH^2+\TH^3\right).
\label{eq:fourth-factor}
\end{equation}
Substitution into Eq.~\eqref{eq:Sgen-basic} yields
\begin{align}
\dot S_{\rm gen}
&=
\frac{\mathcal K(M,t)(\Tb-\TH)^2}{\Tb\TH}
\left(
\Tb^3+\Tb^2\TH+\Tb\TH^2+\TH^3
\right)
\nonumber\\
&\geq0.
\label{eq:Sgen-positive}
\end{align}
The inequality follows term by term for positive temperatures and
\(\mathcal K\geq0\).  Equality holds only at instantaneous temperature
matching or vanishing coupling.

\subsection{Free-energy gradient-flow form}

At fixed bath temperature define
\begin{equation}
F_b(M;\Tb)\equiv M-\Tb S_{\rm res}(M).
\label{eq:Fbath}
\end{equation}
Using \(\dd S/\dd M=1/\TH\),
\begin{equation}
\frac{\partial F_b}{\partial M}
=1-\frac{\Tb}{\TH}
=\frac{\TH-\Tb}{\TH}.
\label{eq:Fprime}
\end{equation}
Introduce the nonnegative mobility
\begin{equation}
\mathcal L(M,t)
\equiv
\mathcal K(M,t)\TH
\left(
\Tb^3+\Tb^2\TH+\Tb\TH^2+\TH^3
\right)\geq0.
\label{eq:mobility}
\end{equation}
It is strictly positive when \(\mathcal K>0\).
Then
\begin{align}
-\mathcal L\frac{\partial F_b}{\partial M}
&=
\mathcal K(\Tb-\TH)
\left(
\Tb^3+\Tb^2\TH+\Tb\TH^2+\TH^3
\right)
\nonumber\\
&=
\mathcal K(\Tb^4-\TH^4)
=\dot M.
\label{eq:gradient-flow}
\end{align}
At fixed \(\Tb\),
\begin{equation}
\frac{\dd F_b}{\dd t}
=\frac{\partial F_b}{\partial M}\dot M
=-\mathcal L
\left(\frac{\partial F_b}{\partial M}\right)^2
\leq0.
\label{eq:Fdecrease}
\end{equation}
The flow therefore moves downhill in free energy.  Because the
Schwarzschild stationary point is a maximum, downhill motion carries the
system away from \(M_\star\): a slightly lighter hole evaporates and a
slightly heavier one accretes.

\subsection{Schwarzschild mass law and unstable thermal threshold}

For a transparent constant-greybody model, write
\begin{equation}
\mathcal K(M,t)=\lambda A_h(M),
\qquad
A_h=16\pi G^2M^2,
\qquad
\lambda>0,
\label{eq:Kmodel}
\end{equation}
where \(\lambda\) absorbs angular capture, polarization, and greybody
information.  This is a thermodynamic transport model, not an exact
replacement for spectral Page emission or relativistic accretion.
Equations \eqref{eq:mass-balance} and \eqref{eq:sch-temp} give
\begin{align}
\dot M
&=16\pi\lambda G^2M^2
\left[
\Tb^4-\left(\frac{1}{8\pi GM}\right)^4
\right]
\nonumber\\
&=16\pi\lambda G^2M^2\Tb^4
-\frac{\lambda}{256\pi^3G^2M^2}.
\label{eq:masslaw-sch}
\end{align}
Define
\begin{equation}
\alpha\equiv\frac{\lambda}{256\pi^3}>0.
\label{eq:alpha-def}
\end{equation}
Then the vacuum limit is the familiar inverse-square mass law
\begin{equation}
\dot M=-\frac{\alpha}{G^2M^2},
\label{eq:vacuum-mdot}
\end{equation}
consistent in structure with detailed Hawking flux calculations
\cite{Page1976,MacGibbon1991}.

At fixed \(\Tb\), the zero of Eq.~\eqref{eq:masslaw-sch} is precisely
\begin{equation}
M=M_\star=\frac{1}{8\pi G\Tb}.
\label{eq:separatrix}
\end{equation}
Let \(f(M)=\mathcal K(M)(\Tb^4-\TH^4)\).  At \(M_\star\), the factor in
parentheses vanishes, so the derivative of \(\mathcal K\) does not
contribute:
\begin{align}
f'(M_\star)
&=
\mathcal K_\star
\left[-4\TH^3\frac{\dd\TH}{\dd M}\right]_{M_\star}.
\label{eq:fprime1}
\end{align}
For Schwarzschild,
\begin{equation}
\frac{\dd\TH}{\dd M}=-\frac{\TH}{M},
\label{eq:THprime}
\end{equation}
and \(\TH(M_\star)=\Tb\), hence
\begin{equation}
f'(M_\star)
=\frac{4\mathcal K_\star\Tb^4}{M_\star}>0.
\label{eq:fprime-positive}
\end{equation}
Writing \(M=M_\star+\Delta M\), the linearized equation is
\begin{equation}
\dot{\Delta M}
=\frac{4\mathcal K_\star\Tb^4}{M_\star}\Delta M
+\Order(\Delta M^2).
\label{eq:linear-instability}
\end{equation}
The instantaneous temperature-matching mass is therefore linearly
unstable.

In an expanding radiation universe, Eq.~\eqref{eq:Mstar-t} implies
\begin{equation}
\dot M_\star=\frac{M_\star}{2t}>0.
\label{eq:Mstar-moving}
\end{equation}
For \(\Delta(t)=M(t)-M_\star(t)\),
\begin{equation}
\dot\Delta
=f(M,t)-\frac{M_\star}{2t}.
\label{eq:moving-delta}
\end{equation}
Even a PBH exactly at \(M_\star\) at one instant has
\(\dot\Delta=-M_\star/(2t)<0\) if its mass does not immediately grow.  Cosmic
cooling produces a moving zero-flux threshold, not an invariant separatrix
or a persistent equilibrium trajectory.

\subsection{Two analytic limiting solutions}

\paragraph{Evaporation-dominated limit.}
Integrating Eq.~\eqref{eq:vacuum-mdot},
\begin{align}
M^2\dd M&=-\frac{\alpha}{G^2}\dd t,
\nonumber\\
\int_{M_i}^{M(t)}m^2\dd m
&=-\frac{\alpha}{G^2}\int_{t_i}^{t}\dd t',
\nonumber\\
\frac{M^3(t)-M_i^3}{3}
&=-\frac{\alpha}{G^2}(t-t_i),
\nonumber\\
M(t)
&=\left[
M_i^3-\frac{3\alpha}{G^2}(t-t_i)
\right]^{1/3}.
\label{eq:evap-solution}
\end{align}
Within the semiclassical extrapolation, the evaporation time is
\begin{equation}
\tau_{\rm evap}
=\frac{G^2M_i^3}{3\alpha}.
\label{eq:evap-time}
\end{equation}
The last Planckian stage lies outside the assumptions of
Sec.~\ref{sec:assumptions}.

\paragraph{Accretion-dominated radiation limit.}
Neglecting the evaporation term and substituting
Eq.~\eqref{eq:Trad4} into the first term of
Eq.~\eqref{eq:masslaw-sch},
\begin{align}
\dot M
&\simeq
16\pi\lambda G^2M^2
\frac{45}{16\pi^3Gg_\ast t^2}
\nonumber\\
&=
\frac{45\lambda G}{\pi^2g_\ast}
\frac{M^2}{t^2}.
\label{eq:accretion-approx}
\end{align}
Define
\begin{equation}
a_\lambda\equiv\frac{45\lambda}{\pi^2g_\ast}.
\label{eq:alambda}
\end{equation}
Then
\begin{equation}
\frac{\dd}{\dd t}\left(\frac{1}{M}\right)
=-\frac{\dot M}{M^2}
=-\frac{a_\lambda G}{t^2}.
\label{eq:inverseM}
\end{equation}
Integrating,
\begin{align}
\frac{1}{M(t)}-\frac{1}{M_i}
&=-a_\lambda G
\int_{t_i}^{t}\frac{\dd t'}{t'^2}
\nonumber\\
&=a_\lambda G
\left(\frac{1}{t}-\frac{1}{t_i}\right),
\label{eq:accretion-integrated}
\end{align}
or
\begin{equation}
M(t)
=
\left[
\frac{1}{M_i}
+a_\lambda G
\left(\frac{1}{t}-\frac{1}{t_i}\right)
\right]^{-1}.
\label{eq:accretion-solution}
\end{equation}
Writing \(\chi=a_\lambda GM_i/t_i\), the asymptotic mass is finite when
the limiting denominator is strictly positive, namely \(\chi<1\), in which
case \(M_\infty=M_i/(1-\chi)\).  At \(\chi=1\), the denominator is
positive at every finite time but \(M(t)=t/(a_\lambda G)\) has no finite
asymptotic limit.  For \(\chi>1\), a finite-time denominator zero marks
the breakdown of the fixed-background and constant-\(\lambda\)
approximations, not a physical divergence.  A complete
treatment would solve relativistic fluid accretion and cosmological
backreaction, extending the Bondi and Michel frameworks
\cite{Bondi1952,Michel1972}.  The limited radiation-era growth is consistent
with the classic PBH analysis of Carr and Hawking
\cite{CarrHawking1974}.

\section{General analytic deformation of the horizon residue}
\label{sec:deformation}

\subsection{Perturbative residue ansatz}

Let a corrected but still nonextremal one-parameter horizon have
\begin{equation}
\Rh(M)
=2GM\left[1+\epsilon\psi(M)\right],
\qquad
|\epsilon\psi(M)|\ll1,
\label{eq:Rdeform}
\end{equation}
where \(\epsilon\) is a bookkeeping parameter and \(\psi\) is dimensionless.
This ansatz is meaningful only after the corrected field equations,
solution branch, horizon, and asymptotic time normalization have been
specified.  Residue corrections in modified-gravity temperature
calculations motivate such a parametrization
\cite{ChenCJP2025}, but Eq.~\eqref{eq:Rdeform} is not itself a field
equation.

The inverse temperature is
\begin{equation}
\beta_h(M)
=4\pi\Rh(M)
=8\pi GM[1+\epsilon\psi(M)].
\label{eq:beta-deform}
\end{equation}
Using \((1+x)^{-1}=1-x+\Order(x^2)\),
\begin{equation}
T_h(M)
=\frac{1}{8\pi GM}
\left[
1-\epsilon\psi(M)
\right]
+\Order(\epsilon^2).
\label{eq:Tdeform}
\end{equation}

Choose a fixed reference mass \(M_{\rm ref}>0\) in the domain of validity,
independent of \(M\), \(\beta_b\), and \(\epsilon\), and write
\[
S_{\rm ref}(\epsilon)
=4\pi GM_{\rm ref}^2+\epsilon C_{\rm ref}+\Order(\epsilon^2).
\]
Here \(C_{\rm ref}\) is independent of \(M\) and must be fixed by an
independent entropy normalization relative to the background.  Assume
\(\psi\) is smooth and the weak-deformation expansion is controlled over
the integration interval.  The entropy follows from
Eq.~\eqref{eq:res-entropy-general}:
\begin{align}
S_{\rm res}(M)
&=S_{\rm ref}(\epsilon)+4\pi\int_{M_{\rm ref}}^M
2Gm[1+\epsilon\psi(m)]\dd m
\nonumber\\
&=4\pi GM^2
+\epsilon C_{\rm ref}
+8\pi G\epsilon
\int_{M_{\rm ref}}^Mm\psi(m)\dd m
+\Order(\epsilon^2).
\label{eq:Sdeform}
\end{align}
Thus the residue fixes the entropy derivative, whereas the reference
entropy is an additional input.  A zero lower limit is justified only if
a controlled, integrable extension to \(M=0\) exists and the normalization
\(S(0;\epsilon)=0\) is imposed; in that limit \(C_{\rm ref}=0\).

\subsection{Shift of the thermal saddle}

At fixed \(\beta_b\), stationarity requires
\begin{equation}
\beta_b
=8\pi GM_\star
\left[1+\epsilon\psi(M_\star)\right].
\label{eq:saddle-deform-eq}
\end{equation}
Write
\begin{equation}
M_\star=M_0+\epsilon\,\delta M+\Order(\epsilon^2),
\qquad
M_0\equiv\frac{\beta_b}{8\pi G}.
\label{eq:Mexpand}
\end{equation}
Substitution into Eq.~\eqref{eq:saddle-deform-eq} gives
\begin{align}
\beta_b
&=8\pi G(M_0+\epsilon\delta M)
\left[1+\epsilon\psi(M_0)\right]
+\Order(\epsilon^2)
\nonumber\\
&=8\pi GM_0
+8\pi G\epsilon
\left[
\delta M+M_0\psi(M_0)
\right]
+\Order(\epsilon^2).
\label{eq:saddle-expand}
\end{align}
Since \(8\pi GM_0=\beta_b\), the coefficient of \(\epsilon\) must vanish:
\begin{equation}
\delta M=-M_0\psi(M_0).
\label{eq:deltaM}
\end{equation}
Therefore,
\begin{equation}
M_\star
=M_0\left[1-\epsilon\psi(M_0)\right]
+\Order(\epsilon^2).
\label{eq:Mstar-deform}
\end{equation}

\subsection{Shift of the nucleation exponent}

The deformed exponent is
\begin{equation}
\mathcal B(M;\beta_b)
=\beta_bM-4\pi GM^2
-\epsilon C_{\rm ref}
-8\pi G\epsilon\int_{M_{\rm ref}}^Mm\psi(m)\dd m
+\Order(\epsilon^2).
\label{eq:Bdeform}
\end{equation}
Evaluate it at \(M_\star=M_0+\epsilon\delta M\).  Taylor expansion gives
\begin{align}
\mathcal B_\star
&=
\underbrace{\left(\beta_bM_0-4\pi GM_0^2\right)}
_{\mathcal B_{\star,0}}
\nonumber\\
&\quad
+\epsilon\delta M
\underbrace{\left(\beta_b-8\pi GM_0\right)}_{0}
\nonumber\\
&\quad
-\epsilon C_{\rm ref}
-8\pi G\epsilon
\int_{M_{\rm ref}}^{M_0}m\psi(m)\dd m
+\Order(\epsilon^2).
\label{eq:Bstar-expansion}
\end{align}
The explicit mass-shift terms cancel because the zeroth-order solution is
stationary.  Thus
\begin{equation}
\boxed{
\mathcal B_\star
=\frac{\beta_b^2}{16\pi G}
-\epsilon C_{\rm ref}
-8\pi G\epsilon
\int_{M_{\rm ref}}^{M_0}m\psi(m)\dd m
+\Order(\epsilon^2).
}
\label{eq:Bstar-deform-final}
\end{equation}
The corresponding logarithmic change in the leading thermal rate is
\begin{equation}
\Delta\ln\Gamma_{\rm th}
=\epsilon C_{\rm ref}+8\pi G\epsilon
\int_{M_{\rm ref}}^{M_0}m\psi(m)\dd m
+\Order(\epsilon^2),
\label{eq:rate-shift}
\end{equation}
apart from changes in the prefactor.  The absolute exponent shift therefore
requires \(C_{\rm ref}\) as well as the residue deformation.

The saddle curvature becomes
\begin{align}
\mathcal B''(M_\star)
&=-4\pi\Rh'(M_\star)
\nonumber\\
&=-8\pi G
\left\{
1+\epsilon
\left[
\psi(M_\star)+M_\star\psi'(M_\star)
\right]
\right\}
+\Order(\epsilon^2).
\label{eq:curvature-deform}
\end{align}
If, in addition to \(|\epsilon\psi|\ll1\), the derivative combination
\(\left|\epsilon[\psi+M\psi']\right|\ll1\) is controlled in the saddle
neighborhood, the Schwarzschild negative mode is unchanged.  Small
amplitude alone does not ensure this derivative bound.
Stabilization would require
\begin{equation}
1+\epsilon[\psi(M_\star)+M_\star\psi'(M_\star)]<0,
\label{eq:stability-condition}
\end{equation}
which is generically nonperturbative or demands a rapidly varying
\(\psi\).

\subsection{Explicit power-law diagnostic}

As a transparent analytic test, take
\begin{equation}
\psi(M)=\psi_0\left(\frac{M}{M_c}\right)^p,
\qquad
p>-2.
\label{eq:psi-power}
\end{equation}
The weak-deformation assumption is imposed on the positive-mass interval
between \(M_{\rm ref}\) and \(M_0\).  Then
\begin{align}
\int_{M_{\rm ref}}^{M_0}m\psi(m)\dd m
&=
\frac{\psi_0}{M_c^p}
\int_{M_{\rm ref}}^{M_0}m^{p+1}\dd m
\nonumber\\
&=
\frac{\psi_0}{p+2}
\frac{M_0^{p+2}-M_{\rm ref}^{p+2}}{M_c^p}.
\label{eq:power-integral}
\end{align}
Equations \eqref{eq:Mstar-deform} and
\eqref{eq:Bstar-deform-final} become
\begin{align}
M_\star
&=
M_0
\left[
1-\epsilon\psi_0
\left(\frac{M_0}{M_c}\right)^p
\right]
+\Order(\epsilon^2),
\label{eq:Mpower}\\
\mathcal B_\star
&=
\frac{\beta_b^2}{16\pi G}
-\epsilon C_{\rm ref}
-\frac{8\pi G\epsilon\psi_0}{p+2}
\frac{M_0^{p+2}-M_{\rm ref}^{p+2}}{M_c^p}
+\Order(\epsilon^2).
\label{eq:Bpower}
\end{align}
For \(-2<p<0\), a formal zero-reference integral converges, but the
deformation is not uniformly small near zero; the positive reference mass
must therefore be retained unless an additional controlled extension is
supplied.
This example is a falsifiable parametrization of analytic structure, not a
claim that a particular quantum-gravity theory selects
\((\psi_0,M_c,p)\).

\section{Independent checks}
\label{sec:checks}

\subsection{Reissner--Nordstr\"om fixed-charge check}

Although the target PBHs are neutral, a charged solution provides a useful
independent algebraic test of the residue--entropy relation.  In geometric
charge notation,
\begin{equation}
F(r)
=1-\frac{2GM}{r}+\frac{q^2}{r^2}
=\frac{(r-r_+)(r-r_-)}{r^2},
\label{eq:RNF}
\end{equation}
where
\begin{equation}
r_\pm=GM\pm\sqrt{G^2M^2-q^2},
\qquad
r_++r_-=2GM,
\qquad
r_+r_-=q^2.
\label{eq:RNroots}
\end{equation}
At the outer horizon,
\begin{align}
\mathcal R_+
&=\Res_{r=r_+}\frac{\dd r}{F(r)}
\nonumber\\
&=\lim_{r\to r_+}
\frac{(r-r_+)r^2}{(r-r_+)(r-r_-)}
\nonumber\\
&=\frac{r_+^2}{r_+-r_-}.
\label{eq:RNres}
\end{align}
Thus
\begin{equation}
T_+
=\frac{r_+-r_-}{4\pi r_+^2},
\qquad
\beta_+
=\frac{4\pi r_+^2}{r_+-r_-}.
\label{eq:RNtemp}
\end{equation}

At fixed \(q\), differentiate the horizon equation
\begin{equation}
r_+^2-2GMr_++q^2=0
\label{eq:RNhorizon-eq}
\end{equation}
with respect to \(M\):
\begin{align}
2r_+\frac{\dd r_+}{\dd M}
-2Gr_+
-2GM\frac{\dd r_+}{\dd M}
&=0,
\nonumber\\
2(r_+-GM)\frac{\dd r_+}{\dd M}
&=2Gr_+.
\label{eq:RNdiff1}
\end{align}
Because \(2(r_+-GM)=r_+-r_-\),
\begin{equation}
\frac{\dd r_+}{\dd M}
=\frac{2Gr_+}{r_+-r_-}.
\label{eq:drplus}
\end{equation}
For \(S_+=\pi r_+^2/G\),
\begin{align}
\left(\frac{\dd S_+}{\dd M}\right)_q
&=
\frac{2\pi r_+}{G}
\frac{\dd r_+}{\dd M}
\nonumber\\
&=
\frac{2\pi r_+}{G}
\frac{2Gr_+}{r_+-r_-}
\nonumber\\
&=
\frac{4\pi r_+^2}{r_+-r_-}
=\beta_+.
\label{eq:RNfirstlaw}
\end{align}
The temperature obtained from the pole and the entropy obtained from the
area law satisfy the fixed-charge first law exactly.

\subsection{Dimension, limit, and sign checks}

The checks supporting the central results are summarized in
Table~\ref{tab:checks}.

\begin{table}[htbp]
\centering
\caption{Research-quality checks for the central results.}
\label{tab:checks}
\begin{tabularx}{\textwidth}{@{}p{0.25\textwidth}p{0.13\textwidth}X@{}}
\toprule
Check & Status & Evidence \\
\midrule
Schwarzschild temperature
& Pass
& Equation \eqref{eq:sch-temp} agrees with surface gravity and Hawking's
result.\\

Schwarzschild entropy
& Pass
& Residue integration gives \(4\pi GM^2=A_h/(4G)\) in
Eqs.~\eqref{eq:S-sch}--\eqref{eq:area-law}.\\

Charged-family first law
& Pass
& Equation \eqref{eq:RNfirstlaw} gives
\((\partial S/\partial M)_q=\beta_+\) exactly.\\

Canonical instability
& Pass
& \(\mathcal B''=-8\pi G<0\), consistent with negative heat capacity and
the Euclidean negative mode.\\

Generalized second law
& Pass
& Equation \eqref{eq:Sgen-positive} is a product of nonnegative factors.\\

Radiation-era dimensions
& Pass
& \(G\Trad^2\) is dimensionless and
\(\mathcal B_\star\propto t/\sqrt G\) is dimensionless.\\

Classical limit of deformation
& Pass
& Setting \(\epsilon\to0\) in
Eqs.~\eqref{eq:Tdeform}, \eqref{eq:Sdeform},
\eqref{eq:Mstar-deform}, and \eqref{eq:Bstar-deform-final} recovers
Schwarzschild.\\

Formation-era adiabaticity
& Pending
& For \(M_i\sim\gamma_{\rm eff}\MH\),
\(2GM_iH\sim\gamma_{\rm eff}\) may be only moderately small; a fully
cosmological metric is required at horizon formation.\\

Modified-gravity entropy
& Pending
& An explicit corrected action and solution must show that
\(\dd S_{\rm Wald}/\dd M=4\pi\Rh\) at fixed couplings.\\

Planck endpoint
& Outside scope
& The semiclassical hierarchy \(M\gg\mpl\) fails.\\
\bottomrule
\end{tabularx}
\end{table}

Independent numerical spot checks were also performed for generic positive
values of \((G,M,\Tb,g_\ast,\lambda)\).  They reproduced
Eqs.~\eqref{eq:sch-temp}, \eqref{eq:dSdM-res},
\eqref{eq:Bstar}, \eqref{eq:Bstar-t}, and
\eqref{eq:RNfirstlaw}; they also gave
\(\dot S_{\rm gen}>0\) on both sides of \(M_\star\).

\section{Falsifiability, limitations, and physical interpretation}
\label{sec:limits}

\paragraph{First falsifying test: independent temperature.}
For any proposed PBH metric, calculate the surface-gravity or
Hamilton--Jacobi temperature independently.  If it differs from
\(1/(4\pi\Rh)\) for the same normalized time and the same horizon, the local
residue prescription fails for that geometry.

\paragraph{Second falsifying test: independent entropy.}
In a corrected theory, compute the Wald entropy and all work terms.  The
one-parameter thermodynamicization used here requires
\begin{equation}
\left(\frac{\partial S_{\rm Wald}}{\partial M}\right)_{\lambda_i,J,Q}
=4\pi\Rh(M).
\label{eq:wald-test}
\end{equation}
Failure of Eq.~\eqref{eq:wald-test} means that the residue does not by itself
generate the entropy; a missing work term, coupling variation, or incorrect
horizon branch must be identified.

\paragraph{Third falsifying test: exchange kernel.}
The sign theorem \eqref{eq:Sgen-positive} assumes detailed balance.  A
microscopic spectral calculation should test whether the net flux can be
written as a positive-kernel integral,
\begin{equation}
\dot M
=\sum_s\int_0^\infty
\dd\omega\,
\Gamma_s(\omega,M)
\omega
\left[
n_s(\omega,\Tb)-n_s(\omega,\TH)
\right],
\qquad
\Gamma_s\geq0.
\label{eq:spectral-kernel}
\end{equation}
Equation \eqref{eq:mass-balance} is the greybody-averaged limit of such a
relation.  Strong nonequilibrium distortions of \(n_s\) invalidate the
fourth-power model but not necessarily the spectral second-law statement.

\paragraph{Local versus cosmological horizons.}
The PBH horizon and the FRW apparent horizon are distinct.  Near formation,
\(r_hH\) need not be parametrically tiny, and a globally static Euclidean
section does not exist.  The residue temperature is then interpreted only
as a local, quasi-stationary scale.  A McVittie-like or numerical
collapse spacetime is needed to quantify corrections of order \(r_hH\).

\paragraph{Thermal nucleation versus perturbative collapse.}
The main robust negative result is that the equilibrium thermal channel is
exponentially suppressed whenever its Schwarzschild saddle is
semiclassical.  This does not suppress PBHs produced by a large primordial
curvature perturbation, a phase transition, bubble collision, cosmic
defect, or an early matter-dominated epoch.  Those mechanisms provide
nonequilibrium initial data and require their own abundance calculations.

\paragraph{Meaning of a residue correction.}
The function \(\psi(M)\) in Eq.~\eqref{eq:Rdeform} is observable only after
time normalization and solution branches are fixed.  A coordinate artifact
or a freely chosen integration constant is not a physical correction.
The reference entropy must also be independently normalized relative to
the background.  After this normalization is fixed, a correction that
survives these checks shifts both the instantaneous thermal threshold and
the exponential nucleation weight
through Eqs.~\eqref{eq:Mstar-deform} and
\eqref{eq:Bstar-deform-final}.  This shared shift is the specific
difference point predicted by the framework.

\section{Conclusions}

We have developed a self-contained thermodynamicization map for
quasi-stationary primordial black holes.  The first certain anchor is the
simple horizon pole of
\(\omega_h=\dd r/\sqrt{AB}\).  Euclidean regularity and the residue theorem
give
\(\beta_h=4\pi\Rh\).  The fixed-work-variable first law then promotes the
same residue to an entropy derivative,
\(\dd S/\dd M=4\pi\Rh\).

For Schwarzschild PBHs, this chain exactly recovers
\[
\Rh=2GM,\qquad
\TH=\frac{1}{8\pi GM},\qquad
\SBH=4\pi GM^2.
\]
The associated canonical exponent has an unstable stationary point
\[
M_\star=\frac{1}{8\pi G\Tb},
\qquad
\mathcal B_\star=\frac{1}{16\pi G\Tb^2}.
\]
In a radiation-dominated universe,
\[
\mathcal B_\star(t)
=\sqrt{\frac{\pi g_\ast}{720}}\,\frac{t}{\tpl}.
\]
Since \(\mathcal B_\star=4\pi(M_\star/\mpl)^2\), equilibrium thermal PBH
production is exponentially suppressed wherever the semiclassical saddle
is reliable.  Standard PBH formation from overdensities remains controlled
by the collapse threshold and critical scaling, not by this canonical
exponent.

For subsequent PBH--bath exchange, detailed balance gives a manifestly
nonnegative generalized entropy production and a free-energy gradient flow.
The temperature-matching mass is a moving zero-flux threshold; its
fixed-bath stationary point is unstable.
For \(a_\lambda GM_i/t_i<1\), radiation dilution gives finite integrated
accretion in the transparent constant-greybody model, while the vacuum
limit reproduces the
\(M^3\)-scaling of the evaporation lifetime.

Finally, a small analytic change of the simple-pole residue produces linked,
calculable shifts in temperature, entropy, critical mass, and, once the
reference entropy is independently fixed, thermal nucleation exponent.
The next decisive test is not another formal
expansion: it is to select an explicit corrected gravitational action, solve
its PBH horizon consistently in a cosmological background, and verify
Eq.~\eqref{eq:wald-test} against the independent Wald entropy.  Until that
test is passed, the deformation sector should be read as a falsifiable
parametrization rather than a completed quantum-gravity prediction.

\section*{Acknowledgments}
This work was supported by the National Natural Science Foundation of China (General Program) under Project Approval No.~11574089.

\end{document}